\documentclass[aps,floats,twocolumn,epsf,prl,showpacs]{revtex4-1}
\usepackage{graphicx}
\usepackage{epstopdf}
\usepackage{hyperref}
\usepackage{color}

\newcommand{\chim}{\chi_{\bf{Q}}}

\begin{document}

\title{Quantum criticality with a twist -\\ interplay of correlations and Kohn anomalies in three dimensions}

\author{T. Sch\"afer$^a$, A. A. Katanin$^b$, K. Held$^a$, and A. Toschi$^a$}

\affiliation{$^a$Institute of Solid State Physics, TU Wien, 1040 Vienna, Austria}
\affiliation{$^b$ Institute of Metal Physics, 620990, Kovalevskaya str. 18, Ekaterinburg, Russia\\
Ural Federal University, 620002, Mira str.  19, Ekaterinburg, Russia}

\date{ \today }

\begin{abstract}
A general understanding of quantum phase transitions in strongly correlated materials is still lacking. By exploiting a cutting-edge quantum
many-body approach, the dynamical vertex approximation, we make an
important progress, determining the quantum critical properties of the
antiferromagnetic transition in the 
fundamental model for correlated electrons, the Hubbard model in three dimensions. In particular, we
demonstrate that -in contradiction to the
conventional Hertz-Millis-Moriya theory-  its  quantum critical
behavior is driven by the Kohn anomalies of the Fermi surface, even when
electronic correlations become strong.
\end{abstract}

\pacs{71.27.+a, 71.10.Fd, 73.43.Nq}
\maketitle

\let\n=\nu \let\o =\omega \let\s=\sigma

\noindent
{\sl Introduction.}  The underlying quantum mechanical nature of the physical world is often elusive at the macroscopic scale of every-day-life phenomena. In the case of solid state 
physics, the most striking manifestations of its quantum origin are confined to very low temperatures, where thermal fluctuations are frozen. An important exception 
is realized where thermodynamic phase transitions (e.g.~to a magnetic
state) are driven to occur at zero temperature, at a quantum critical
point (QCP)\cite{Loehneysen,Sachdev,Kopp,Sachdev2}:  The corresponding
quantum critical fluctuations become then abruptly visible also at sufficiently high temperatures, representing one of the most exciting subjects in condensed matter physics.
\\
While QCPs are actually found experimentally in the phase-diagrams of
several compounds\cite{Loehneysen}, a general theoretical treatment of
their physics is still lacking. Consequently, the analysis of
experiments often remains based on a mere fitting 
of the exponents controlling the critical behavior at the QCPs,
preventing a general comprehension of the phenomenon. The major challenge, in this respect, is the competition of several equally important physical mechanisms, because, at the QCP, both long-ranged space- and time-fluctuations must be treated on an equal footing. In fact, this is only possible in limiting cases, such as in the perturbative regime, by means of 
Moriya\cite{Moriya}-Dzyaloshinskii-Kondratenko\cite{Dzyalosh} theory and the famous renormalization group (RG) treatment by Hertz\cite{Hertz} and Millis\cite{Millis}.
However, an actual comprehension of the experiments based only on these
theories is highly problematic, for two reasons. First of all, most quantum critical materials are strongly correlated.
This is certainly the case for  the (antiferro)magnetic  quantum critical points (QCPs) of transition metals under pressure, such as  Cr$_{1-x}$V$_{x}$ \cite{CrP1,CrP2,CrV1,CrV2} and heavy fermion compounds under pressure or in a magnetic
field, such as in CeCu$_{\text{6-x}}$Au$_{\text{x}}$ \cite{Schroeder2000} and YbRh$_{\text{2}}$Si$_{\text{2}}$ \cite{Custers2003,Paschen2004}.
It has been established that one effect of strong correlations,
namely the breakdown of the ``large'' Fermi-surface containing both
conduction and $f$-electrons and the associated
local quantum criticality \cite{Si2001,Coleman2005}, may lead to
different critical exponents. Nonetheless, we are still far away
 from identifying the universality classes beyond the conventional Hertz-Millis-Moriya (HMM) theory.\\
Besides electronic correlations,  the physics of
QCPs can also be affected by specific properties
of their Fermi surfaces (FS), such as van Hove singularities, nesting, or
Kohn points. The effects thereof are often of minor importance at
high-$T$, but can be amplified in the low-$T$ regime. 
While van Hove singularities and nesting
require special forms of the electronic spectrum, 
Kohn points are more generic and easily occur in three-dimensional
($3d$) \cite{Kohn59,Rice1970} and two-dimensional ($2d$) systems
\cite{Stern1967,Metzner2012,Metzner2014}.  
Kohn points are defined as the points of the FS  that 
(i)  are connected by the spin-density wave (SDW) vector ${\bf Q}$ and (ii) beyond that have {\sl opposite} Fermi velocities. These points are associated to the textbook 
``Kohn anomalies'' of the susceptibilities \cite{Kohn59,magnetictextbooks}, also called $Q=2k_{F}$ anomalies, which is the momentum where they occur for an isotropic FS. The effect of Kohn anomalies on the phonon dispersion is well known \cite{Kohn59} and the 
breakdown of standard HMM theory has been conjectured \cite{Millis,Loehneysen}.\\
In this paper we make significant progress towards a better understanding of QCPs. We demonstrate that FS features in $3d$ lead to an unexpected universality class of its magnetic QCP, 
which also holds in the non-perturbative regime. 
In principle, the complexity of the competing microscopic mechanisms underlying
a quantum phase transition of correlated electrons calls for a quantum many-body
technique capable of treating  {\sl both}, extended spatial and temporal
fluctuations, {\sl beyond} the weak-coupling, perturbative regime. 
The approach we exploit here is the dynamical vertex approximation
(D$\Gamma$A) \cite{DGA,DGA2,abinitioDGA,RohringerThesis,Held}, which is a diagrammatic extension  \cite{Kusonose,DGA,DF,Multiscale,DB,1PI,DMF2RG,Li,Kitatani,TRILEX,QUADRILEX} of dynamical mean
field theory (DMFT) \cite{DMFTRev,Dinfty} built on its two-particle vertices \cite{vertex,diverg}. It has been already successfully used to study  classical, finite temperature
criticality of strongly correlated systems in $3d$ \cite{DGA_3D,Antipov2014,Hirschmeier2015}, as well as long-range antiferromagnetic (AF) fluctuations and their effect on the electronic self-energy in $2d$ \cite{DGA2,DGA3}. In fact,  D$\Gamma$A builds up non-local corrections at all
 length scales on top of DMFT \cite{vertex}, which in turn captures, in a non-perturbative fashion, all purely local temporal correlations \cite{Dinfty}. Hence,
 per construction, the scheme is particularly suited to the study of quantum critical phenomena. 

The obtained phase diagram 
as a function of doping
displays a progressive suppression of the  N\'eel temperature ($T_N$), a crossover to an
incommensurate SDW-order, and 
eventually the vanishing of the magnetic order at a QCP with $\sim 20\%$ doping.
Upon doping, the critical scaling
properties of the second-order magnetic transition
change abruptly from the ones expected for the universality class of
the $3d$ Heisenberg model, a ``classical'' finite-$T$ phase
transition, to a quantum critical behavior  visible in a
relatively broad funnel-shaped temperature region above the QCP.
Our results unveil the importance of Kohn anomalies
 for the scaling properties of the QCP.  In particular, the  $T$-dependence of the magnetic
susceptibility ($\chim \propto T^{-\gamma}$) at the SDW wave-vector $\bf{Q}$ and of the
correlation length ($\xi \propto T^{-\nu}$) largely deviate from the
typical behavior expected from the HMM theory for AF quantum phase transitions
in $3d$. 

\noindent
{\sl Phase diagram.} We  focus here on the
magnetic transitions in the Hubbard model on a simple cubic
lattice:
\begin{equation}
 H=-t\sum\limits_{<ij>\sigma}{c_{i\sigma}^{\dagger}c_{j\sigma}}+U\sum\limits_{i}{n_{i\uparrow}n_{i\downarrow}}\;,
 \label{eqn:hubbard}
\end{equation}
where $t$ is the hopping amplitude between nearest neighbors, $U$ the local Coulomb interaction, $c_{i\sigma}^{\dagger}$ ($c_{i\sigma}$) creates 
(annihilates) an electron with spin $\sigma=\uparrow,\downarrow$ at site $i$, and $n_{i\sigma}=c_{i\sigma}^{\dagger}c_{i\sigma}$; the average density is $n=\left<n_{i\uparrow}\right>+\left<n_{i\downarrow}\right>$. Hereafter all energies are measured in units of $D=2\sqrt{6}t\equiv 1$, twice the standard
deviation of the non-interacting density of states;  we employ $U=2.0$, for which the highest $T_N$ at half-filling is found in both, DMFT
and D$\Gamma$A \cite{DGA_3D}. We do not consider phase-separation\cite{MF_PD}, charge-ordering\cite{Wu2011,Gunnarsson1989} or disorder\cite{CPA_PD}-induced effects.
\begin{figure}[t!] 
        \centering
                \includegraphics[width=0.48\textwidth,angle=0]{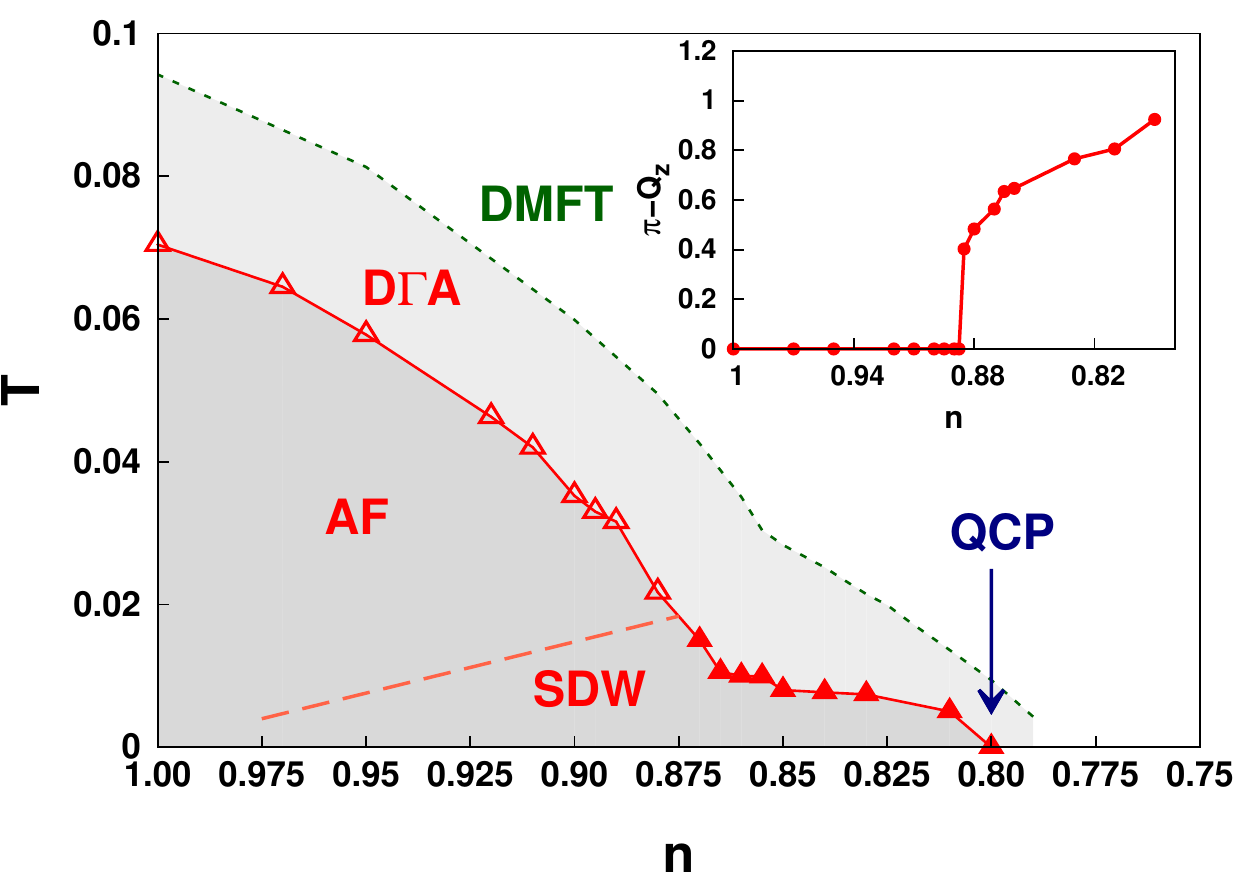}
        \caption{(Color online) \label{Fig1} Phase diagram of the
          $3d$ Hubbard model at $U=2 D$, showing the leading magnetic  instability as a function of the density $n$ in both DMFT
          and D$\Gamma$A.  Inset: Evolution of the magnetic ordering  vector along the instability line of D$\Gamma$A,  showing a transition from an commensurate AF with $Q_{z}\!=\!\pi$ (open triangles in the main panel) to incommensurate SDW with $Q_{z}<\pi$  (full triangles in the main panel). The dashed red line indicates the presumptive crossover between AF and SDW. }
\end{figure}

 To explore the magnetic phase diagram, we employ DMFT with exact diagonalization (ED) as impurity solver 
 and D$\Gamma$A in its ladder-approximation version supplemented by Moriyaesque $\lambda$-corrections, see
 Refs.~\cite{DGA2,RohringerThesis,Rohringer2016} for the
 implementation used here as well [see Supplemental Material, Sec.~II (ii) for more specific details \cite{Suppl}]. This approach includes spin-fluctuations and was successfully applied to calculate the critical
exponents in $3d$ before \cite{DGA_3D}. Superconducting fluctuations are treated at the DMFT level (the full parquet  D$\Gamma$A \cite{Valli2015,Li2016} which would incorporate these fluctuations is numerically too demanding for the required  momentum-grids at the QCP).

The primary quantity we calculate is the static, fully momentum-dependent magnetic susceptibility 
$\chi_{\bf q}\equiv\chi_{\bf q}(\omega\!=\!0)$, as a function of
temperature $T$. It has a maximum at a 
specific (temperature-dependent) wave-vector ${\bf q}={\bf Q}_T$, and diverges at  $T=T_{N}$, marking the occurrence of a second-order phase-transition towards magnetism with  ordering vector ${\bf Q}_{T_N}$. 

Figure \ref{Fig1} shows the corresponding divergence points in the $T$-$n$  phase-diagram both for DMFT (green) and D$\Gamma$A (red).
By progressively reducing $n$, $T_N$ decreases and two regions of the magnetic
ordering can be distinguished: (i) close to
half-filling, we observe an instability at ${\bf Q}_{T_{N}}=(\pi,\pi,\pi)$ 
i.e., to commensurate AF (open triangles); (ii)
at higher doping  ($n\lesssim 0.88$) the ordering vector is
shifted to ${\bf Q}_{T_{N}}=(\pi,\pi,Q_{z}<\pi)$, i.e. an
incommensurate SDW (filled triangles). The
inset of Fig.~\ref{Fig1} quantifies the incommensurability  $\pi-Q_{z}$, i.e.,  the
deviation from a checkerboard AF order.
\begin{figure*}[ht!] 
        \centering
                \includegraphics[width=\textwidth,angle=0]{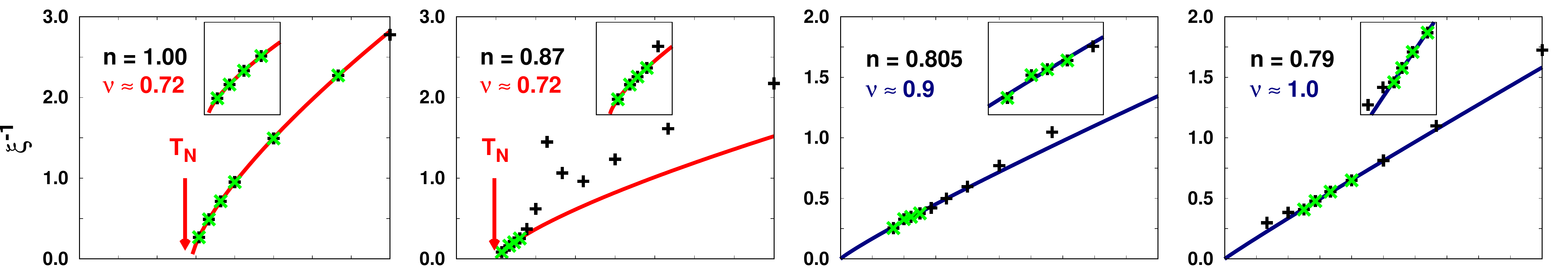}
                \vskip -4.53mm
                \includegraphics[width=\textwidth,angle=0]{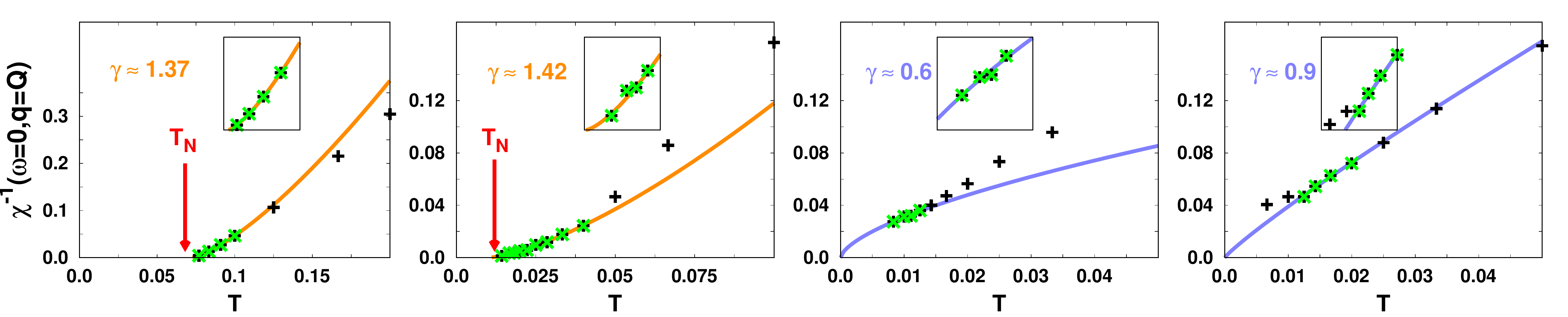}
        \caption{(Color online) \label{Fig3} Inverse correlation
          length ($\xi^{-1}$, upper panels) and maximal susceptibility
          ($\chi^{-1}$, lower panels) computed in D$\Gamma$A as a
          function of $T$ for different $n$. The solid lines show the fits for extracting the critical exponents $\nu$ and $\gamma$ (using the respective green points). The insets show zooms of the four respective lowest temperature points.}
\end{figure*}

Eventually,  ordering is  suppressed completely as
$T_N\rightarrow 0$, leading  to the emergence of a QCP  at $n^{\text{D}\Gamma\text{A}}_{c}\approx{0.805}$. We note that
  the critical filling in DMFT is comparable to that  obtained before \cite{Jarrell} for a similar interaction strength ($U=2.04D$). 

\noindent
{\sl Critical properties}. Let us now turn to the (quantum) critical behavior. We select
representative temperature cuts at four different dopings 
($n=1.0/0.87/0.805/0.79$)  chosen  on both the ordered and the disordered side of the QCP.  Along
these four paths we compute two fundamental
observables, which yield the (quantum) 
critical exponents  $\gamma$ and $\nu$ of the magnetic transition: (i) the 
 spin-susceptibility $\chi_{{\bf Q}_T} \propto (T-T_N)^{-\gamma}$
at its maximum, reached at the  $T$-dependent wave-vector ${\bf Q}_T$,
and (ii)  the corresponding correlation length, $\xi \propto (T-T_N)^{-\nu}$.
The latter is calculated via $\chi_{{\bf Q}_{T}+{\bf q}}=A({\bf q}^2+\xi^{-2})^{-1}$. 

Figure \ref{Fig3} shows the $T$-dependence of $\xi^{-1}$ (upper panels) and $\chi^{-1}_{{\bf Q}_T}$ (lower
panels). Note that, apart from its intrinsic $T$-dependence, the
susceptibility is also affected by the $T$-dependence of the wave
vector ${\bf Q}_T$, with the further complication that the dominating wave-vector changes with both $n$ and $T$.

In the half-filled case (leftmost panels of 
Fig.~\ref{Fig3}) both $\xi$ and $\chi_{{\bf Q}_T}$
display a critical behavior compatible with the $3d$ Heisenberg
universality class when approaching 
the classical (finite-$T$) antiferromagnetic phase transition at
$T_{N}(n=1)\approx{0.072}$. The numerically extracted critical
exponents of $\nu\approx{0.72}$ and $\gamma\approx{1.37}$ are
consistent with previous
calculations \cite{DGA_3D,Hirschmeier2015}, cf.\ our overview in Fig.~\ref{Fig2}  below.

Significant changes  are observed at a doping, where the
SDW-order appears ($n\simeq 0.87$, second column of 
Fig.~\ref{Fig3}). 
Here, by inspecting $\xi^{-1}(T)$ and $\chi^{-1}(T)$, a clear crossover is found
between the high-temperature region ($T>0.04$), where commensurate
AF fluctuations dominate [maximum of
$\chi_{\bf q}$ at $(\pi,\pi,\pi)$], to the low-temperature
regime ($T<0.025$) where incommensurate fluctuations  at $(\pi,\pi,Q_z<\pi)$ outpace these before approaching
the phase-transition. At the crossover,  $\xi^{-1}$ shows a maximum in
Fig.~\ref{Fig3}, which is, however, not an indication of 
a decreasing correlation length, but rather  reflects the inapplicability of our standard definition of $\xi$: 
In the vicinity of the AF-to-SDW crossover, we have a  double-peak structure in $\chi_{\bf q}$ (not shown)
at  $Q_z=\pi$ and   $Q_z\!\sim \! \pi\!-\!0.4$ which altogether appears in the form of a large peak width, i.e., a large  $\xi^{-1}$.

Despite the apparently more complex temperature-behavior of $\xi$ and $\chi$, and the
onset of an incommensurate order, the critical exponents at low $T$
are not altered at all ($\nu\approx{0.72},\gamma\approx{1.42}$) with respect 
to the $3d$ Heisenberg values.
 This is ascribed to the persistence of a classical phase-transition at
$T_{N}(n=0.87)\approx{0.012}$, which still belongs to the same universality
class as the commensurate one.  At higher $T$ a linear behavior of the
inverse susceptibility (which is characteristic for a mean-field theory for bosonic degrees of freedom) is eventually recovered.

{\sl Quantum criticality}. Before turning to our D$\Gamma$A data at the QCP, let us briefly discuss the analytical results for the non-uniform susceptibility in the random phase approximation (RPA). 
We start by recalling that the standard HMM approach relies on the expansion \cite{Loehneysen,Moriya,Dzyalosh,Hertz,Millis}
\begin{equation}
\label{EQ:HMM}
\chi_{\bf{Q+q}} (\omega) =A ({\bf{q}}^2+\xi ^{-2}+i \omega /|{\bf q}|^{z-2})^{-1},
\end{equation}
 where the first and third term in the denominator are determined by
 the band dispersion (under the assumption that no Kohn points
 exist). The $T$-dependence of the correlation length is 
 $\xi^{-1}\propto  T^{\nu}$ with $\nu=(d+z-2)/(2z)=3/4$ ($d=3$ and $z=2$ for a SDW). It originates from the (para)magnon interaction,
 dominating over the $T$-dependence from
 the bare susceptibility. Since $d+z>4$ we are above the upper critical dimension,  and quantum criticality can be described by a bosonic mean-field theory.

 \begin{figure}[bt!] 
        \centering
                \includegraphics[width=0.33\textwidth,angle=0]{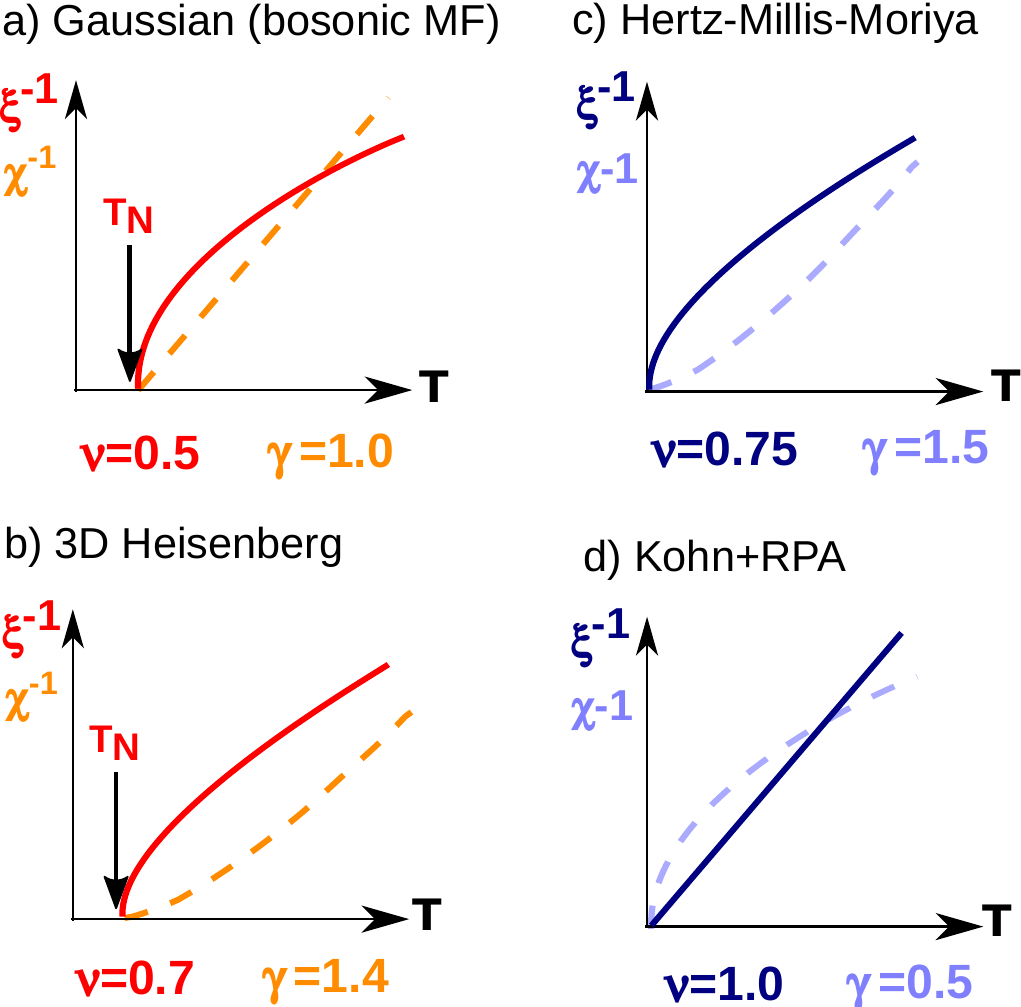}\includegraphics[width=0.164\textwidth,angle=0]{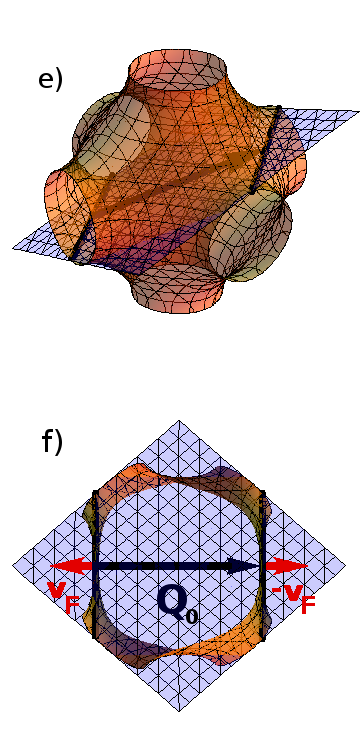}
        \caption{(Color online) \label{Fig2} 
        {\bf (a, b) }
Magnetic correlation length $\xi$ and maximal magnetic susceptibility $\chi$ vs.~$T$ 
 comparing the critical exponents $\nu$ and $\gamma$ for a classical finite-temperature phase transition in
 (a) mean-field theory and (b) for the $3d$ Heisenberg model.
 {\bf  (c, d)}  Quantum critical behavior comparing (c) standard HMM theory and (d) our scenario with Kohn line anomaly.
 {\bf  (e)} Visualization of (one out of four pairs of) Kohn lines in the $3d$ FS of the
          simple cubic lattice with nearest-neighbor hopping and the connecting SDW  vector ${\bf Q}_{0}$. 
{\bf  (f)} $2d$ cut with the Kohn-line of (e) and the corresponding (opposite) Fermi-velocities.}
\end{figure}
 
As shown in the Supplemental Material \cite{Suppl}, for the Kohn points on the FS spin fluctuations are, however, enhanced due to their antiparallel Fermi velocities, and 
their quantum critical behavior changes dramatically.
Moreover, as our D$\Gamma$A calculations below demonstrate, the Kohn quantum critical
behavior survives also in the strongly correlated regime. 
While the (possible) inapplicability of HMM in the presence of  Kohn  points has been pointed out before \cite{Millis,Loehneysen},
their implication on the quantum critical behavior in $3d$ and particularly
the critical exponents have not been analyzed hitherto.

For the simple
cubic lattice, which we consider here for the numerical comparison with the D$\Gamma$A below, there are four pairs of lines of Kohn points
$(\pm K_x,\mp K_x-\pi,-Q_z/2)$ and $(\pi\pm K_x,\mp K_x,Q_z/2)$ 
 which are  connected
by the ground-state spin density  wave-vectors ${\bf Q}_{0}=(\pi,\pi,Q_z)$ (and symmetrically equivalent wave-vectors)  and have
opposite Fermi velocity, see Fig.~\ref{Fig2} e) and f). The leading contributions in the
momentum and $T$-dependence of $\chi^{-1}$ are non-trivial already in RPA. They stem from the vicinity of the lines' endpoints $(0,\pi,\pm Q_{z}/2)$ and $(\pi,0,\pm Q_{z}/2)$, yielding (see Supplemental Material \cite{Suppl}):
\begin{equation}
\chi _{\mathbf{Q}_{T}+{\bf q}}\simeq \left[( \chi^{-1} _{%
\mathbf{Q}_{0}}) _{T=0}+AT^{1/2}+B{T^{-3/2}}q_{z}^{2}\right]^{-1}  \; . \label{Kohn_m_div}
\end{equation}
Here ${\bf Q}_T=\mathbf{Q}_{0}+(0,0,\delta Q_{z})$, with $\delta Q_{z}=-2CT$ describing a shift of the wave-vector with the temperature and $A,B,C$ are positive factors, containing weak, $\ln \ln(1/T)$, corrections.  The susceptibility
Eq.\ (\ref{Kohn_m_div}) is in stark contrast to the standard expansion
 Eq.\ (\ref{EQ:HMM}). It is strongly anisotropic in momentum and  strongly $T$-dependent due to non-analytic momentum- and temperature dependences of the bare susceptibility in the presence of Kohn anomalies. For $q_{z}=0$ we obtain the
critical exponent  $\gamma=1/2$ for the susceptibility, 
whereas the critical  exponent for $\xi$  (defined in the direction of the $z$ axis) is $\nu=1$. 
These exponents are strikingly different from those of HMM theory,
$\nu=3/4, \gamma=2\nu=3/2$. Even their relative magnitude is reversed, see Fig.~\ref{Fig2} c) and d).

A corresponding, radical modification of the critical properties at the QCP (at $n_{\text{c}}=0.805$) is found also
numerically in D$\Gamma$A, see Fig.~\ref{Fig3} (3rd column). Here,
the critical exponents change to $\nu={0.9}$ $(\pm 0.1)$ and
$\gamma={0.6}$ $(\pm 0.1)$ (with an additional error of the same magnitude stemming from the selection of the proper $T$ range, see  Fig.~\ref{Fig3};  a detailed error analysis can be found in  Supplemental
Material \cite{Suppl} Sec.~II). These exponents are in stark contrast
to any standard expectation such as the $3d$ Heisenberg results or HMM theory, but agree with our  RPA exponents. Even when considering the significant error bars,
it is safe to say that only the Kohn-anomaly scenario is consistent with our  D$\Gamma$A results as these irrevocably show
 a roughly   {\sl linear} behavior of $\xi^{-1}(T)$ in the whole
  low- and intermediate $T$-regime above the QCP  (i.e., $\nu\approx 1$) and, even more clear-cut, a {\em strong violation} of the scaling relation $\gamma=
2\nu$ \cite{notevio}, implying a highly non-trivial anomalous dimension $\eta$. 

Slightly overdoping the system (4th column of Fig.~\ref{Fig3}, $n\!=\!0.79$) yields a Fermi-liquid with a finite $\chi$ for $T \rightarrow 0$. In the quantum critical regime (i.e., excluding the low-temperature points which lie outside the quantum critical region) we find
similar exponents as at  optimal doping ($\nu\!\approx\!1.0$, $\gamma\!\approx\!0.9$; the determination of the accurate value of the critical exponent $\gamma$ is more difficult because of the restricted temperature range). 

 No univocal prediction can be made
 instead for the {\sl dynamical} exponent $z$: The frequency dependence of $\chi_{\bf
   q}(\omega)$ in the presence of Kohn anomalies has a
 rather complicated form \cite{Altshuler1995,Metzner2014},  {\sl not}
 characterized by a single exponent. The same effect is also
 responsible for a non-Fermi-liquid power-law in the $2d$ self energy \cite{Metzner2014}. 

Having whole {\sl lines} of Kohn points and hence the above critical exponents  is evidently  specific
 to the $3d$ dispersion with nearest neighbor hopping.
 Consistent with 
 the results of previous studies \cite{Rice1970}, however, we demonstrate in  Sec.~I~D of the Supplemental Material  \cite{Suppl}
 that the critical exponents are $\nu\!=\!\gamma\!=\!1$ for the more general situations of a FS with {\sl isolated} Kohn
points having opposite masses in two directions. This again violates the HMM prediction. Please note that these values of the exponents in $3d$ coincide (up to logarithmic corrections) \cite{note2d} 
   with those expected for Kohn points in $2d$ \cite{Metzner2012}. 
   
   In general, the momentum dependence of vertex corrections beyond RPA and the self-energy corrections should not be too strong, and the quasiparticle damping should be sufficiently small at $T\rightarrow 0$ to preserve the above-mentioned values of the critical exponents in the interacting model. Under these assumptions, we expect the observed behavior to be universal, with several new 'universality classes' depending on whether there are lines of Kohn points with  divergent or non-divergent mass, or isolated Kohn points with opposite masses (see Supplemental Material \cite{Suppl}).

The final outcome of 
our calculations, i.e.\ unusual values of
$\nu$, $\gamma$ and of their mutual relation  which are in a different universality class than HMM theory, can be understood thus 
as the consequence of two competing  physical
processes at work:  On the one hand, as $T_N \rightarrow 0$ at the QCP,
the temporal fluctuations increase the effective
dimension of the system above the three geometrical ones. This 
pushes it above the upper critical dimension and renders  non-Gaussian fluctuations irrelevant as in HMM. On the other hand, the 
effect of Kohn anomalies, yielding a non-analytic momentum- and temperature dependence of the susceptibility, are no longer smeared out by finite $T$ and become relevant.

{\sl Conclusions.}  We have studied the magnetic QCP in the doped $3d$ Hubbard model.
We find that, even above the upper critical dimension, quantum
criticality is {\sl not} of standard Hertz-Millis-Moriya type. 
Even in the presence of strong correlations critical properties are driven by Fermi-surface features: 
the presence of Kohn points leads
to unexpected critical exponents, the breakdown of the scaling relations
and not univocal definitions of the dynamical exponent $z$.
The implications of our results go well beyond the specific system
considered and also hold for other dispersion relations, showing how strongly
the QCP physics can be driven by peculiar features of the FS.
In this perspective, the cases where controversial interpretations of
 experiments in the proximity of QCPs arise might need to be reconsidered. 

\acknowledgments{{\sl Acknowledgments.} We thank G. Rohringer, A. Eberlein, W. Metzner and S. Paschen for insightful discussions and  dedicate our 
 work in memory of Walter Kohn and his multitudinous, groundbreaking contributions to solid state physics.
We acknowledge support from the Austrian Science Fund
(FWF) through the Doctoral School ``Building Solids for Function'' (TS, FWF project W1243), the project I-610 (TS, AT) and project I 1395-N26 as part of the DFG research
unit FOR 1346 (KH), as well as from the 
European Research Council under the European Union's Seventh Framework Program (FP/2007-2013)/ERC grant agreement n.\ 306447 (KH). The  work of AK is performed within the theme "Electron" 01201463326 of FASO, Russian Federation and Russian Foundation for Basic Research grant No 17-02-00942. Calculations were performed on the Vienna Scientific Cluster (VSC); we thank J. Zabloudil and M. St{\"o}hr for the great support.}

\pagestyle{empty}

\clearpage

\noindent
\hskip -12mm
\includegraphics[width=1.10\textwidth]{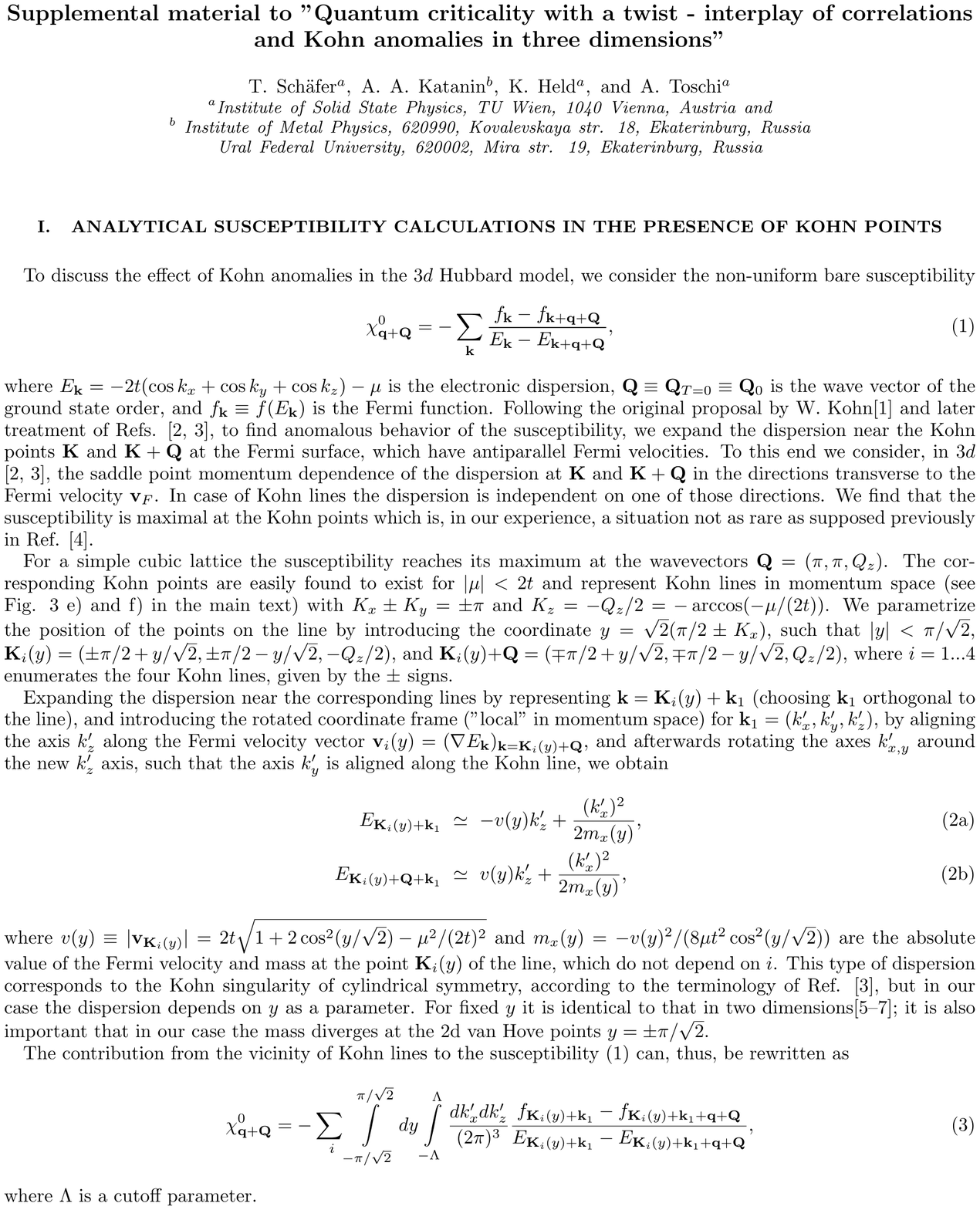}

\clearpage

\hskip -12mm
\includegraphics[width=1.10\textwidth]{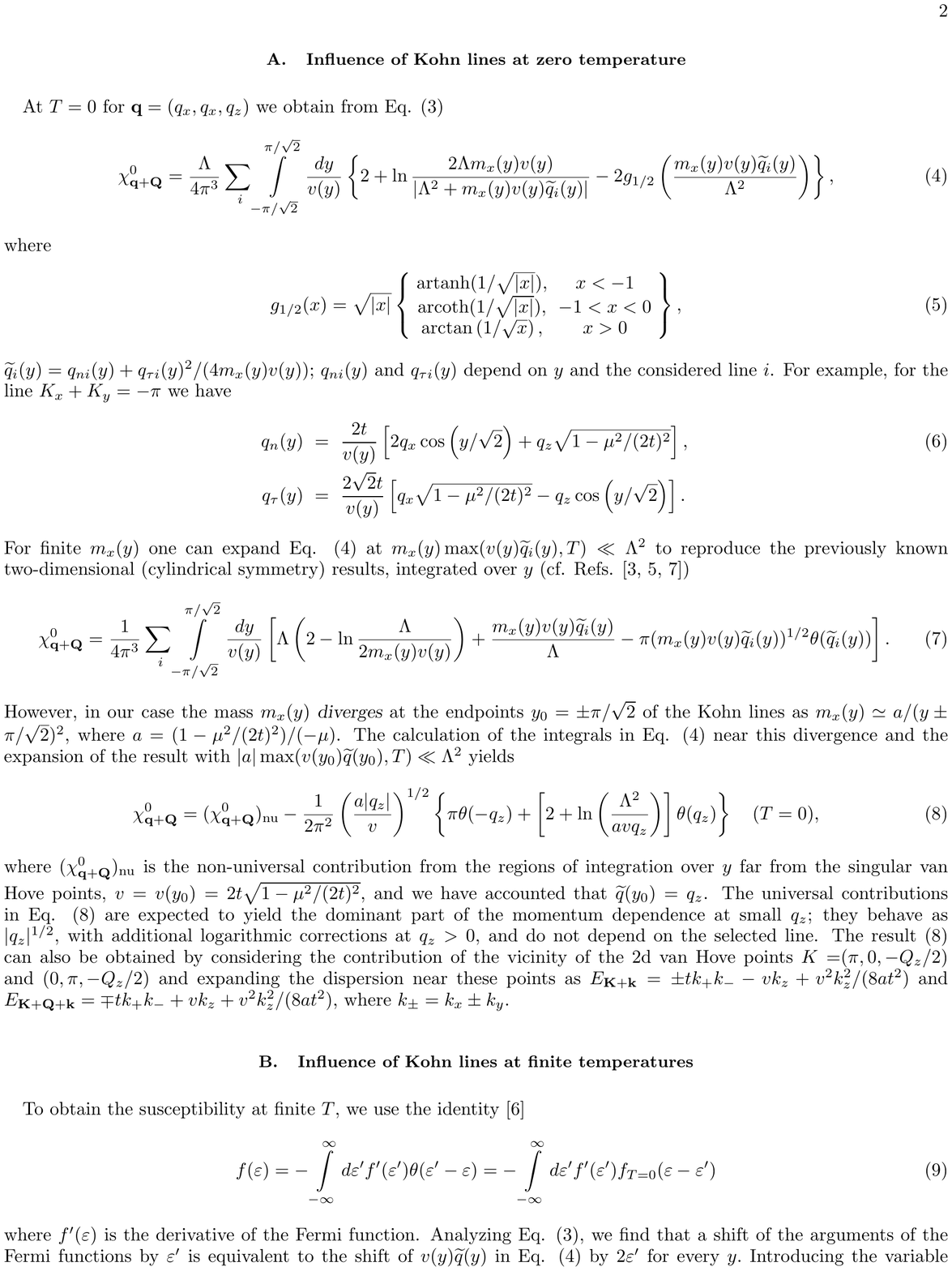}

\clearpage

\hskip -12mm
\includegraphics[width=1.10\textwidth]{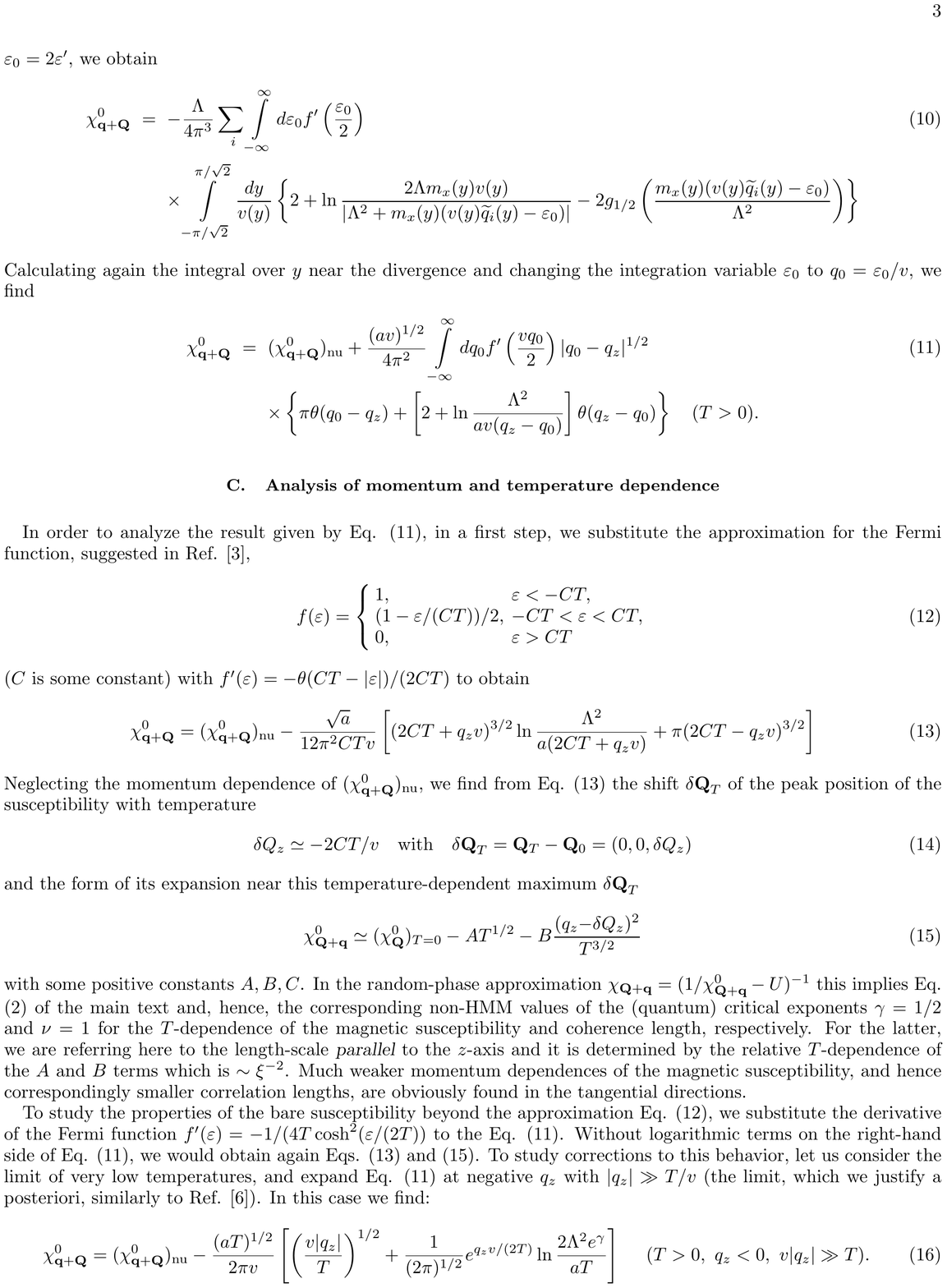}

\clearpage

\hskip -12mm
\includegraphics[width=1.10\textwidth]{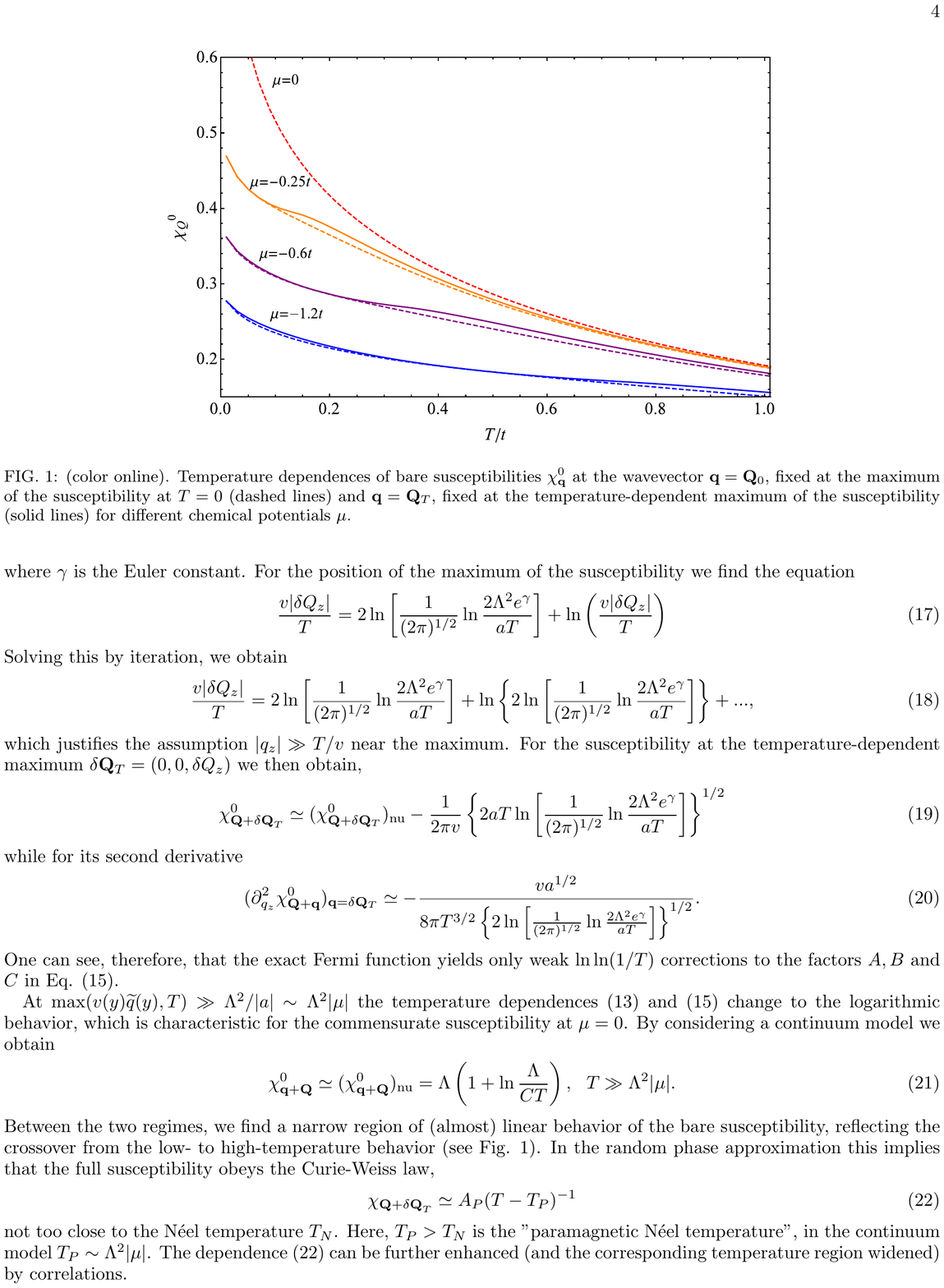}

\clearpage

\hskip -12mm
\includegraphics[width=1.10\textwidth]{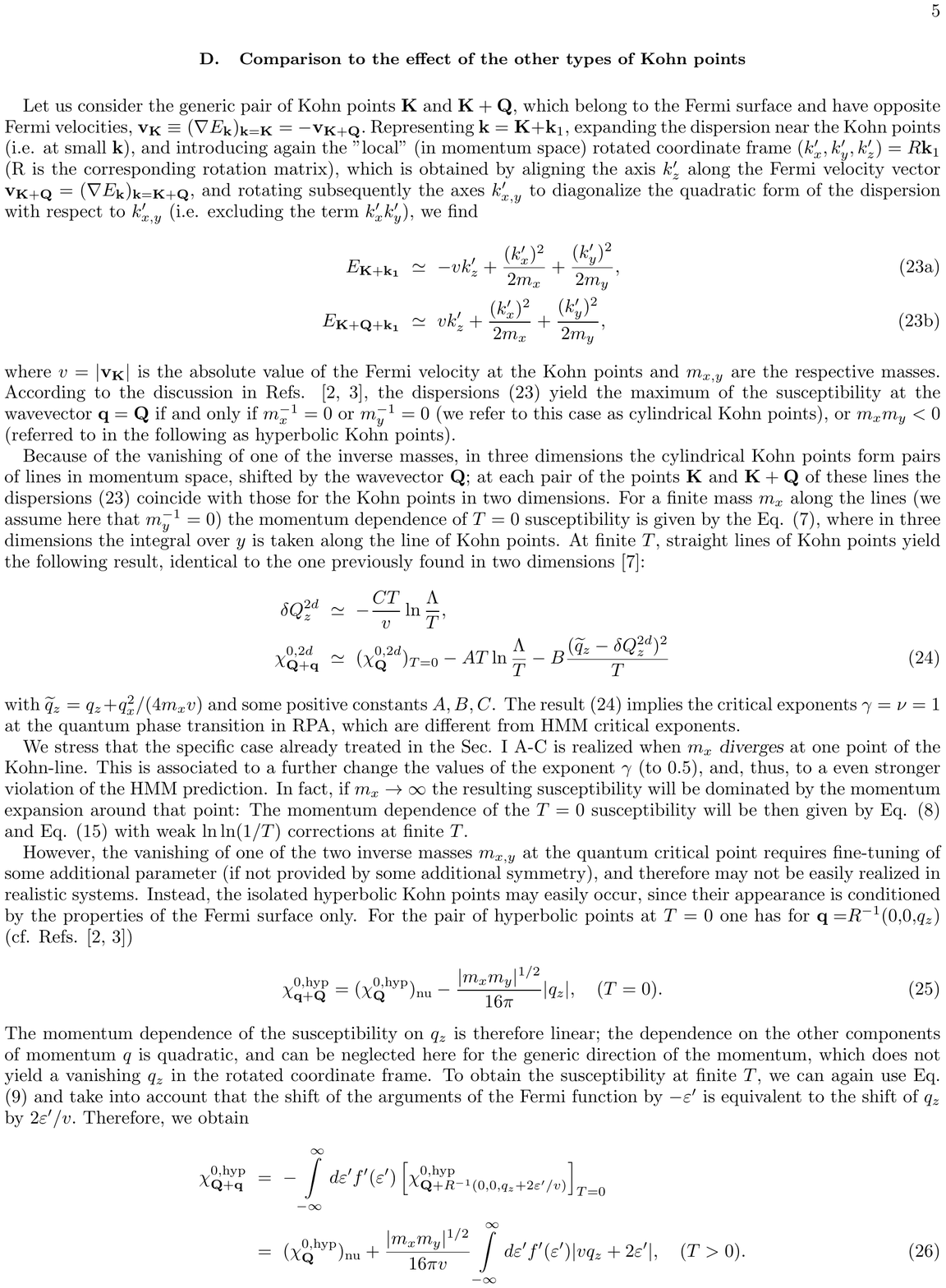}

\clearpage

\hskip -12mm
\includegraphics[width=1.10\textwidth]{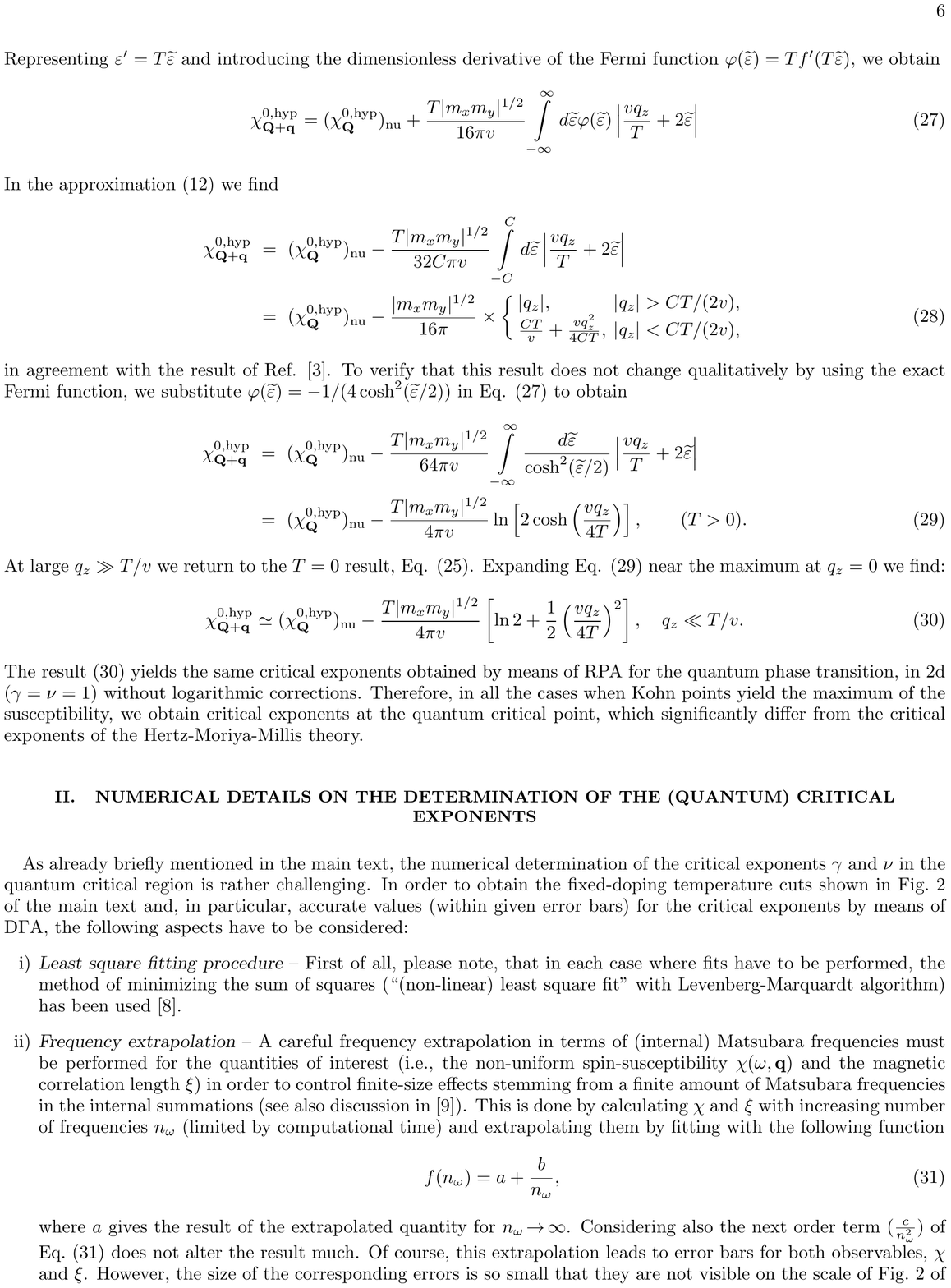}

\clearpage

\hskip -12mm
\includegraphics[width=1.10\textwidth]{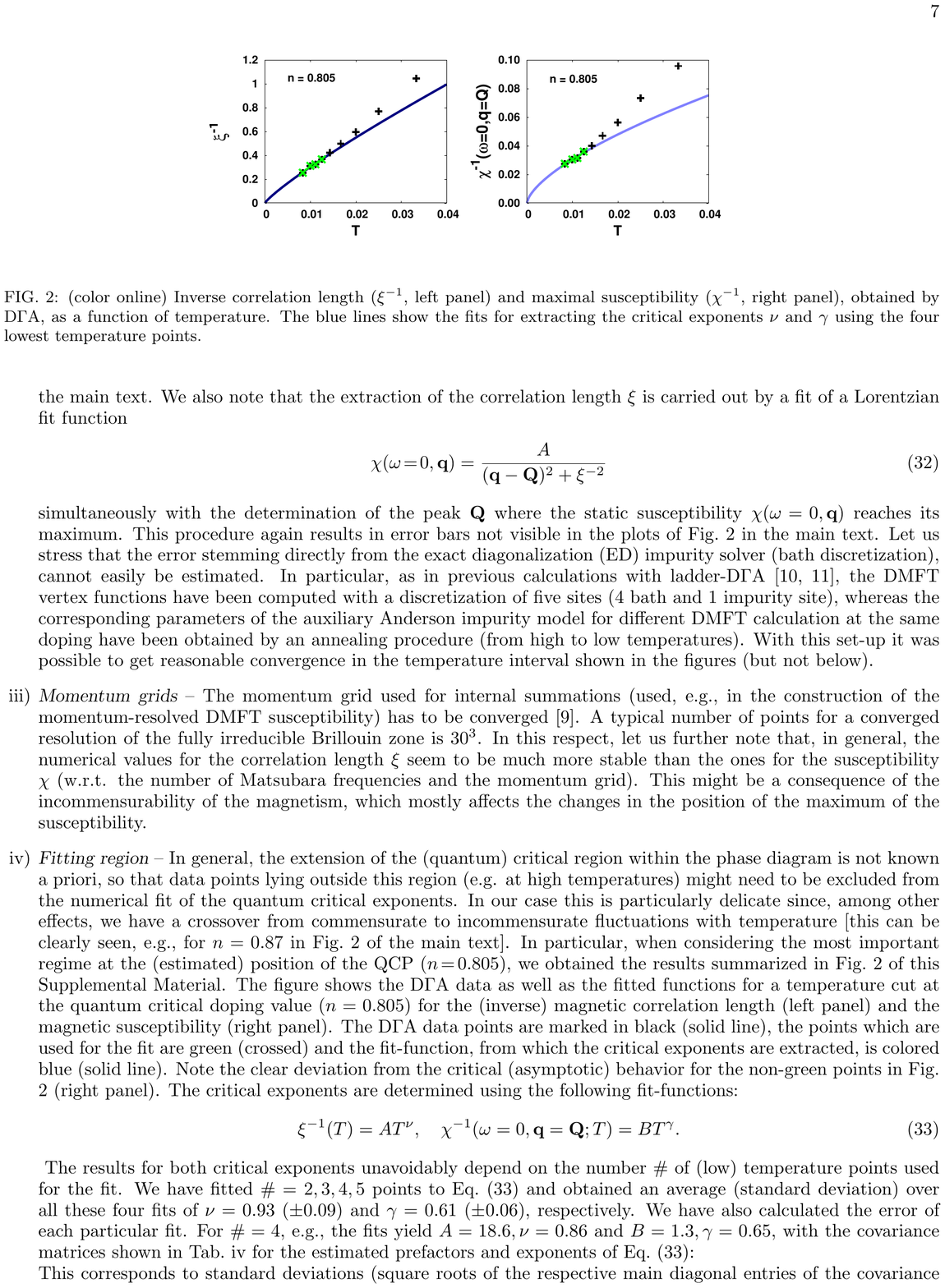}

\clearpage

\hskip -12mm
\includegraphics[width=1.10\textwidth]{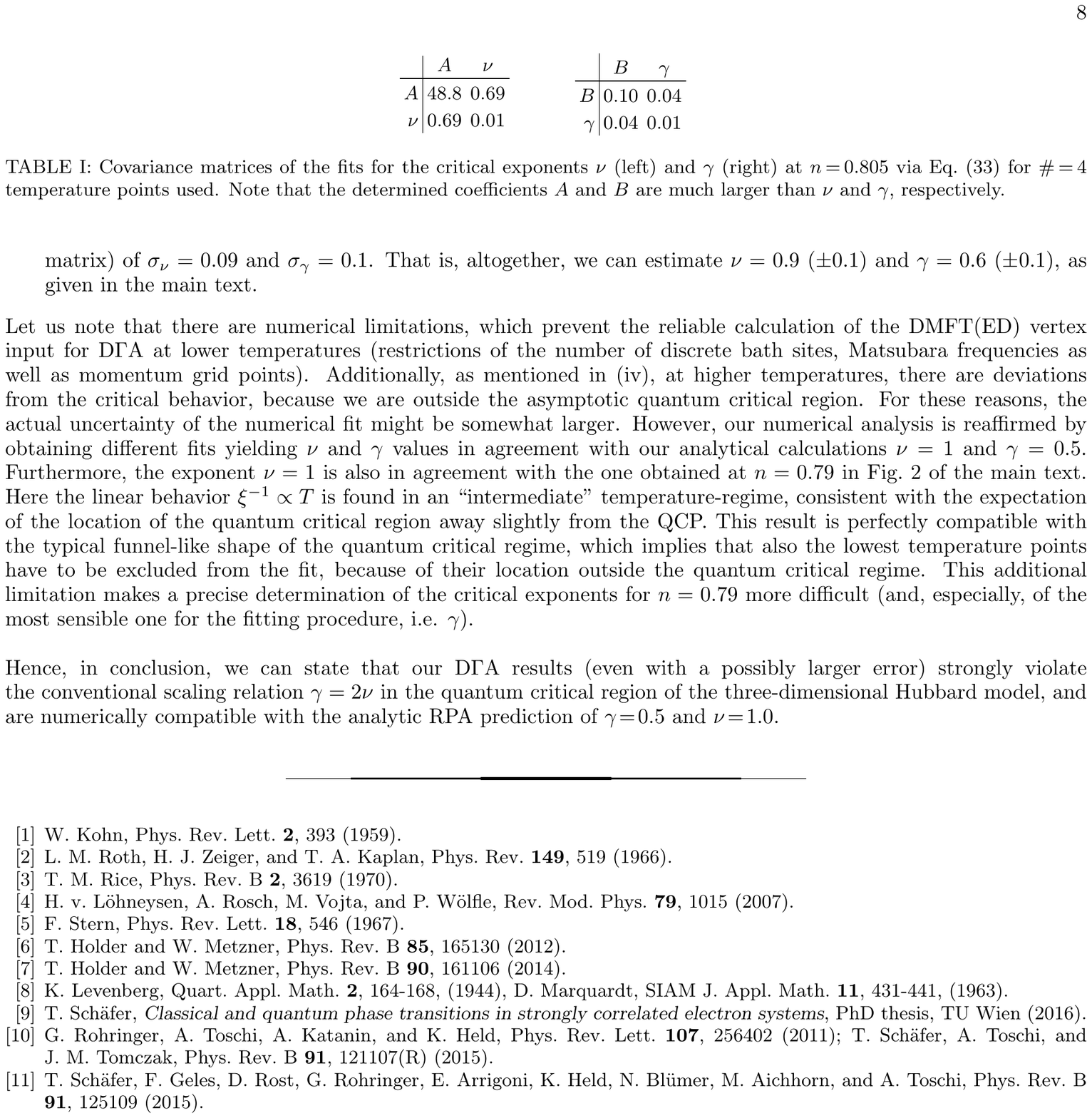}

\clearpage

\end{document}